\documentclass[11pt]{article}
\usepackage{geometry}                
\usepackage{graphicx}
\usepackage{amssymb}
\usepackage{epstopdf}
\title{Space and time from translation symmetry}
\author{A. Schwarz}

\begin{document}

\maketitle
\newcommand{\ket}{\rangle}
\newcommand{\bra}{\langle}
\newcommand{\ba}{{\mathbf a}}

\newcommand{\la}{{a_{\text{L}}}}
\newcommand{\ra}{{a_{\text{R}}}}
\newcommand{\ta}{\tilde{a}}
\newcommand{\cA}{{\mathcal A}}
\newcommand{\bA}{{\mathbf A}}
\newcommand{\bB}{{\mathbf B}}
\newcommand{\cB}{{\mathcal B}}
\newcommand{\bC}{{\mathbb C}}
\newcommand{\cC}{{\mathcal C}}
\newcommand{\cF}{{\mathcal F}}
\newcommand{\tF}{{\tilde{F}}}
\newcommand{\bH}{{\bf H}}
\newcommand{\cH}{{\mathcal H}}
\newcommand{\tH}{\tilde{H}}
\newcommand{\tJ}{\tilde{J}}
\newcommand{\cL}{{\mathcal L}}
\newcommand{\tl}{{\tilde{l}}}
\newcommand{\bp}{{\bf p}}
\newcommand{\bP}{{\bf P}}
\newcommand{\bx}{{\bf x}}
\newcommand{\bR}{{\mathbb R}}
\newcommand{\res}{\text{res}}
\newcommand{\tS}{\tilde{S}}
\newcommand{\cT}{{\mathcal T}}
\newcommand{\tw}{\tilde{w}}
\newcommand{\tz}{\tilde{z}}
\newcommand{\bZ}{{\mathbb Z}}
\newcommand{\bX}{{\mathbf X}}


 {\bf Abstract}
 
 We show that the notions of space and time in algebraic quantum field theory arise from translation symmetry if we assume asymptotic commutativity. We argue that this construction can be applied to string theory.
 \vskip .2in

The goal of the present paper is to understand better the foundations of string theory.

 I will start with  the question "What is quantum field theory?" I  hope to explain that in quantum field theory space and time can be considered as secondary notions arising from the existence of commutative symmetry group of the theory. The explanation will be based on the paper [1].  This paper generalizes Haag -Ruelle theory [3] [4 ] of scattering  in algebraic quantum field theory {\footnote {It was suggested that more appropriate name for axiomatic quantum field theory in its modern form is algebraic quantum field theory [18]. I agree with this suggestion. }}to the case when the assumption of locality or of commutativity of observables in domains separated by a space-like interval is replaced by the assumption of asymptotic commutativity. {\footnote {In the case of local theory this condition is satisfied for quasilocal observables. }} (See, for example, [2] for  a review of Haag-Ruelle  theory.)  Haag-Ruelle constructions as well as the results of [1] can be applied only if all particles are massive. However,  D. Buchholz and his collaborators developed scattering theory also in the case when there exist massless particles (see [5], [6],[7]). One can show that in this case also one can get rid of the assumption of locality replacing it by the condition of asymptotic commutativity in a form much weaker than in [1]. It seems that the condition of asymptotic commutativity is sufficient to guarantee macroscopic causality.

 It seems that the explanation of origin of space-time from translation symmetry   can be applied also to string theory. At the end of the paper we 
 will discuss the application of our ideas to string field theory. The string theory is non-local, however, people understand that still it should be local in some sense (see, for example, [17]).
Our statement is that  asymptotic commutativity
replaces locality in string theory. We argue that
that the problems of string theory can be rigorously formulated in the framework  of algebraic  quantum field theory.

\section {What is quantum mechanics?}

Let us begin with the formulation of quantum mechanics in terms of the algebra of observables. (The material of this section is standard; see, for example, [2].)

The starting point of this formulation is a unital associative algebra $\cal A$  over  $\bC$ (the algebra of observables ). 
One assumes that  this algebra is equipped with antilinear involution  $A\rightarrow A^*$. One says that a linear functional $\omega$ on $\cA$ specifies a state if $\omega (1)=1 $ and $\omega (A^*A) \geq 0$ (i.e. if the functional is normalized and positive).  The probability distribution $\rho (\lambda)$ of real observable $A=A^*$ in the state $\omega$  is defined by the formula $\omega (A^n)=\int \lambda ^n \rho (\lambda) d\lambda$.

In the textbooks on quantum mechanics the algebra of observables
consists of operators acting on a (pre)Hilbert space . Every vector $x$ having a unit norm  specifies a state by the formula 
$\omega (A)= \bra Ax,x\ket$. ( More generally, a density matrix $K$
defines a state $\omega (A)={\rm Tr}  AK$.) This situation is in some sense universal: for every state $\omega$ on $\cA$ one 
construct a (pre)Hilbert space $\cH$ and a representation of $\cA$ by operators on this space in such a way that the state $\omega$ corresponds to a vector in this space. 

To construct       $\cH$ one defines inner product  on $\cA$ by the formula 
$\bra A,B \ket=\omega (A^*B)$. The space $\cH$ can be obtained from $\cA$ by means of factorization with respect to zero vectors
of this inner product. The inner product on $\cA$ descends to
$\cH$ providing it with a structure of preHilbert space. The state
$\omega $ is represented by a vector of $\cH$  that corresponds to the unit  element of $\cA$.  An operator of multiplication  from the left
by an element $C\in \cA$ descends to an operator $\hat C$ acting on $\cH;$ this construction gives a representation of $\cA$ in the algebra of bounded operators on $\cH$. The
algebra of operators of the form $\hat C$ where
$C\in \cA$ will be denoted by   $\hat {\cA}$.

It is important to notice that although every state of the   algebra $\cA$ can be represented by a vector in Hilbert space in general one cannot  identify Hilbert spaces corresponding to different states.

Time evolution in algebraic formulation is specified by one-parameter group $\alpha (t)$ of automorphisms 
  of the  algebra $\cA$ preserving the involution. This group acts in obvious way on the space of states. If $\omega$ is a stationary state ( a state
  invariant with respect to time evolution ) then the
  group $\alpha (t)$ descends to   a group $U(t)$ of unitary transformations of corresponding space $\cH$. The generator $H$ of $U(t)$ plays the role of Hamiltonian. If the spectrum of $H$ is 
  non-negative one says that the stationary state $\omega$ is a ground state.
  
  \section {What is quantum field theory?}
  
  Quantum field theory can be regarded as a particular case of quantum mechanics .
  
  Let us consider quantum field theory on $d$-dimensional space
  (on $(d+1)$-dimensional space -time). Then it is natural to assume that the group of translations of space-time acts on the algebra of observables.  
Therefore we can say that quantum field theory is quantum mechanics with action of commutative Lie group on the algebra of observables  $\cA$ .  In other words, we assume that operators
$\alpha (\bx,t)$ where $\bx \in \bR ^d$, $t\in \bR$ are automorphisms of $\cA$ preserving the involution  and that
$\alpha (\bx,t) \alpha (\bx ^{\prime},t ^{\prime})=
\alpha (\bx +\bx ^{\prime}, t+t ^{\prime})$. We will use the notation
$A(\bx,t)$ for $\alpha (\bx, t) A$ where $A \in \cA$. 

 Let us consider now a ground state $\omega$  that  is invariant with respect to translation group. (The action of translation group on $\cA$ induces the action of this group on the space of states.)
Then  the action of translation group on $\cA$ generates unitary 
representation of this group on  (pre)Hilbert space $\cH$. Generators of this representation $\bP$ and$-H$ are identified with momentum operator and Hamiltonian.  The state $\omega$  is considered as physical vacuum; it has zero momentum and energy. The corresponding vector in the space $\cH$ will be denoted by $\Phi$. We define one-particle state as a generalized $\cH$-valued function $\Phi (\bp)$ obeying $\bP \Phi (\bp)=\bp \Phi (\bp),
H \Phi (\bp)=\varepsilon (\bp) \Phi (\bp)$. (More precisely, for some
class of test functions $f(\bp)$ we should have a linear map 
$f\rightarrow \Phi (f)$ of this class into $\cH$ obeying
$\bP \Phi (f)= \Phi (\bp f), H\Phi (f)=\Phi (\varepsilon (\bp) f)$.) Instead of notation $\Phi (\bp)$ we will use later the notation $|\bp>.$

If we know only $\bP$ and $H$ we can define a notion of one-particle state, but we cannot analyze the scattering of states belonging to the space $\cH$ (the scattering of excitations of ground state ). However, we can define scattering matrix  if we impose the condition of asymptotic commutativity on the representation of the algebra of observables $\cA   $ (asymptotic commutativity of the operator algebra $\hat {\cA}$). More precisely, we assume that  
$$\parallel [\hat {A}(\bx,t),\hat {B}]\parallel \leq C \frac {1+|t|^s}{1+|\bx|^n}.$$
Here $\hat {A}\in \hat {\cA}$ and $\hat {B}\in \hat {\cA}$ are operators on $\cH$ representing elements $A \in \cA$,$B \in \cA$, $n$ is arbitrary, $C$ and $s$ depend on $n$.

Assuming asymptotic commutativity of $\hat {\cA}$ one can   prove (under certain conditions) that the scattering matrix does exist. 
{\footnote { In addition to asymptotic commutativity one should assume that the stability of one-particle states is guaranteed by conservation laws. One assumes that there exists a gap between   zero energy of the vacuum and the remaining part of energy spectrum. One should impose also some restrictions on  one-particle dispersion laws $\varepsilon _i (\bp) $;  for example, it is sufficient to require that these functions are strictly convex. }}This is 
the main result of [1] . It follows from this result that an observer interested in  evolution of states  represented by vectors belonging to the space $\cH$ can interpret the events he considers as happening in $(d+1)$-dimensional space-time.
 
Notice that we do not need Lorentz-invariance
to define the scattering matrix and to prove its existence. 

  The  results of [1]  cannot be applied to the case when there exist massless particles. In this case one should relax the condition of asymptotic commutativity and modify the notion of particle.  The scattering theory in the presence of massless particles was developed in important papers by Buchholz and his collaborators   in the framework of Haag-Araki
axiomatics ( [5]-[7] ; see also [2] or [10] for review and [8] for detailed treatment). However, locality is not very relevant in this theory; 
it is possible to generalize the theory to the case of  algebras obeying  asymptotic commutativity condition in much weaker form than in [1]. Namely, one should require that the representation of the algebra $\cA$ corresponding to the physical vacuum obeys
\begin{equation}
\label{ac}
\int  || [\hat {A}(\bx,t),\hat {B}] ||d\bx <c(t)
\end{equation}
for every $A\in \hat{\cA}, B\in\hat{\cA}$.  We  assume that $c(t)$ grows at most polynomially
as $t\to \infty.$

  Adding to this requirement  some conditions on energy-momentum spectrum one can prove all results of [5]  ; it  is sufficient to assume that the energy-momentum spectrum lies in a cone $K$ located in the  half-space where the energy is positive. {\footnote {The paper [5] contained
a statement that one can get rid of the locality assumption in the formulation of main results, however, the conditions necessary for validity of these results were not stated. }} In relativistic theory this condition is satisfied: the energy-momentum spectrum lies in the forward lightcone $E\geqslant ||\bp ||$. Without loss of generality we assume that $K$ coincides with the forward lightcone.  The construction of collision cross-sections given in [6],[7]
can be repeated in  this more general situation.

Let us describe the main ideas of this construction. First of all we notice that one can
extend the algebra $\hat {\cA}$ consisting of
operators $\hat A$ where $A\in \cA$  to a  algebra $\cB$  that is invariant with respect to Hermitian conjugation, obeys the asymptotic commutativity condition \ref {ac} and contains
together with every operator $B$ all operators
of the form $B(f)=\int f(\bx,t)B(\bx,t)d\bx dt$ where
$f$ is rapidly decreasing function. (This follows from the remark that adding $B(f)$ to an asymptotically commutative algebra containing $B$ we do not spoil asymptotic commutativity.)
For every operator $B\in \cB$ we construct
an operator $L\in \cB$ as an operator $B(f)$ where $f$ is a function whose  Fourier transform $\tilde {f}$
has support in compact set in  the complement of the cone $K$. The operator $L$ annihilates the
vacuum  vector $\Phi$. It follows from our assumptions that  the distance of the support of $\tilde {f}$ from the cone $K$ is positive therefore we can assume that for some $\delta >0$ the support of $\tilde {f}$ lies in the subset of the energy-momentum space specified by the condition $E<|\bp|-\delta.$  In this case $L$ annihilates also all states having energy less than $\delta$. To check this fact we notice that the matrix element
$<a'|B(f)|a>$ where $a$ and $a'$ are (not necessarily normalizable) eigenvectors of energy-momentum operator with eigenvalues
$(E,\bp)$ and $(E',\bp ')$ vanishes unless the
vector $(E'-E,\bp '-\bp)$ belongs to the support of
$\tilde {f}$.  

If $G$ is a domain in energy-momentum space we denote by $S(G)$ the set of  elements of $\cH$ that can be represented as linear combinations of eigenvectors of energy-momentum operator with eigenvalues $(E, \bp)\in G$ (in other words, $S(G)$ is a set of vectors with energy-momentum in the domain $G$). The above remark can be interpreted as a statement that
\begin{equation}
\label{sg}
B(f)S(G)\subset S(G+{\rm supp}\tilde {f}).
\end{equation}

The set of operators  $L$ constructed above will be denoted by $\mathcal{L}$. 

We will see that the operator $C=L^*L$ can be considered as a detector counting particles with energies $>\delta$. More precisely, for such an operator we can construct a family  of operators
\begin{equation}
\label{c}
C_t(h)=\int h(\frac{\bx}{t},) C(\bx,t)d\bx.
\end{equation} 
The following estimate can be derived from our assumptions
\begin{equation}
\label{e}
||C_t(h) P(E)||\leq \gamma ||h||_{\infty}
\end{equation}
where  $\gamma$ does not depend on $t$ and $h$. Here $P(E)$ stands for the projector on the
space of states having energy $\leq E$ and
$||h||_{\infty}={\rm sup}|h(\bx)|$. One can reformulate (\ref{e}) as a statement that for
any normalized vector $\Psi$ having energy
less than $E$ we have
$$||C_t(h)\Psi ||\leq \gamma ||h||_{\infty}.$$

For any state $\Psi$ having finite energy
we consider expressions of the form
\begin{equation}
\label{t}
<\Psi, C^{(1)}_t(h^{(1)})\cdots C^{(n)}_t(h^{(n)}) \Psi>
\end{equation}
where $C^{(i)}=(L^{(i)})^*L^{(i)}$ are operators of the kind considered above and $h^{(i)}$ are bounded functions.
These expressions are bounded functions of $t$
therefore it is natural to expect that they have a finite limit as $t$ tends to $\pm \infty$. If the theory has a particle interpretation one can
express these limits in terms of 
momentum densities of particles [11]. {\footnote { The paper [11] is based on Haag-Ruelle scattering theory; in particular, it assumes Lorentz invariance. However, the same considerations go through in the assumptions of [1].}} Namely, if  for
$t\to +\infty$ the state $\Psi$ can be considered as a superposition of multi-particle states with momentum space densities $\rho ^{out}_n(\bp _1,...,\bp _n)$  we obtain  that in this limit (\ref {t})  becomes
\begin{equation}
\label{tt}
\int d\bp _1 ...d\bp _n \rho^{out} _n(\bp _1,...,\bp _n)\prod _kh^{(k)}(\frac{\partial \varepsilon (\bp)}{\partial \bp}|_{\bp=\bp _k})<\bp_k|C^{(k)}|\bp _k>.
\end{equation}
Here $<\bp |C|\bp '>$ stands for the matrix element of $C$ between one-particle states $|\bp>=\Phi (\bp)$; we restrict this expression to the diagonal $\bp=\bp'.$ (One can prove that the matrix element is non-singular on the diagonal.)
For $t\to -\infty$ we obtain similar relation
with momentum space densities $\rho ^{in}_n$.

If we know collision cross sections we can express $\rho ^{out}$ in terms of $\rho ^{in}$.
Conversely, we can determine collision cross sections if we know $\rho ^{in}$ and $\rho ^{out}$ for sufficiently large class of vectors $\Psi$, operators $C^{(i)}$ and functions $h^{(i)}$.

We have assumed that there exists only one kind of particles  without internal degrees 
of freedom; the generalization to the case when this assumption is not satisfied is trivial (one should include summation over discrete indices into above formula.)

To explain the role of asymptotic commutativity in the above statements we will formulate a crucial lemma from [5] :

{\bf Lemma} Let us suppose that the operator
$A$ acting on $\cH$ obeys $A(\bx _1)\cdots A(\bx  _n)P(E)=0$ for all $\bx _i\in \mathbb{R}^d.$ Then
\begin{equation}
\label{b}
||P(E)\int _D d\bx A^*(\bx)A(\bx)P(E)||\leq  (n-1)\int  d\bx ||[A(\bx), A^*]||.
\end{equation}

Here $D$ (the domain of integration in LHS) is an arbitrary compact subset. If $A\in \cB$ the integral in RHS converges due to (\ref {ac}). If $ A=L\in \mathcal{L}$ is one of operators used in the construction of detectors the conditions of Lemma are satisfied  for sufficiently large $n$. 
This fact follows  from (\ref {sg}).

The notion of particle we were using is not always appropriate. In particular, due to the "photon
cloud" surrounding an electron it cannot be applied in quantum electrodynamics. One can generalize the notion of particle considering positive translationally invariant
linear functionals on the space of detectors .
(Such functionals are called particle weights.) 
We define here the space of detectors as a linear space spanned by operators of the form $C=L^*L$ where $L\in \cL$; the translations act on this space in natural way. A particle specifies a particle 
weight as $<\bp|C|\bp>$. The particle weights form a cone; it is sufficient to consider only extremal points of this cone.

Let us consider again a limit of (\ref {t}) assuming that it has the form (\ref {tt}) where
$<\bp_k|C^{(k)}|\bp _k>$ are particle weights.
Then we can regard (\ref {tt}) as a definition of $\rho ^{out}$ ( as a definition of $\rho ^{in}$ if $t\to -\infty$). Again using this definition we define collision cross-sections.

\section {Generalizations and comments}

  1. One can generalize the above setup to include fermionic
 particles.  In this generalization one assumes that  the algebra
 $\cA$ is $ \bZ _2$-graded ; the commutator should be replaced
 by supercommutator.

 2. If the action of translation group on algebra $\cA$ can be extended to  the action of Poincare group and the ground state $\omega$
is invariant with respect to this action then we can define an action of Poincare group on the  corresponding space $\cH$. If the assumption of asymptotic commutativity of representation of $\cA$ on $\cH$ is satisfied and we can construct the scattering matrix  this matrix
is Lorentz-invariant.

It is very difficult to construct non-trivial examples of relativistic theories obeying the asymptotic commutativity assumption.
One can hope that every renormalizable local   Lorentz-invariant action functional gives such example.  (This hope is based on the fact that one  perform  the construction
in the framework of perturbation theory.)

 3. We considered the case when the translation group is continuous. One can  define the notion of asymptotically 
commutative algebra also in the situation when the
translation group is discrete and construct the scattering
matrix in this situation. Some results of this kind were 
proved by Barata [9] in the framework of lattice models.

 4. In the consideration of gauge theories it is useful to utilize BRST formalism. This means that one should work with  a differential algebra $\cA$ and modify all definitions in appropriate way.
For example,  we can restrict ourselves to physical observables and physical states (i.e. to observables and states annihilated by the differential). 
In the condition of asymptotic commutativity we consider only physical observables; the commutator should become small after
the addition of a homologically trivial observable.  

Notice, that the use of BRST formalism is very natural from the viewpoint of mathematics.  In homological
algebra it is very useful to replace a module with its resolution, i.e. to consider a differential module that is quasiisomorphic to the original one. Similarly, one can consider instead of an associative algebra a quasiisomorphic differential algebra.
In BRST formalism we replace a complicated algebra of observables with simpler algebra   equipped with a differential.
(For example, we can remove constraints  extending the algebra
and introducing a differential.)

 5. One can relax conditions imposed on the ground state  requiring invariance only with respect to a subgroup of translation  group. Then the dimension of the space-time for 
corresponding observer will be smaller. 

The ground state is not necessarily unique. It is   possible that  different ground states lead to different dimensions of 
space-time .   Let us emphasize that in our construction the notion of space-time does exist for an observer interested in
finite-energy excitations of a ground state and depends on the choice of the ground state.
  
  6.  We considered scattering theory for excitations of ground
  state. One can consider also scattering of excitations
of an equilibrium state; in this case one should  talk about quasiparticles  (or thermal particles) instead of particles. More generally it is possible to
study locally equilibrium states and their excitations. Notice that in this case we have local notions of space and time, but  global notions of space and
time  do not necessarily  exist. 

Let us stress that the  notion of local equilibrium can be defined without reference to space-time picture. The main step is the definition of locally stationary state. (In stationary state the
correlation functions $\omega (A(t))$ do not depend on $t$; in locally stationary state we require  these functions to be slowly
varying. Similarly, instead of translation invariance of the state
$\omega$ we assume that $\omega (A(\bx,t))$ varies slowly.)  
  
\section{What about string theory?}

Let us consider superstring field theory in 10-dimensional flat space.
(One can talk about any form of this theory).
I would like to argue that it is possible that the origin of space-time in superstring theory can be explained on the basis of the considerations of Sec 2 . (It is important to notice that the results
of [1] are not sufficient in this case due to existence of massless particles ; therefore it is necessary to apply the constructions of [5],[6],[7] that were explained above.) We can talk also about bosonic string field theory in 26-dimensional flat space, but in this case we should  disregard the presence of tachyon in the spectrum. (It is a standard practice  to  pretend that there exists
no tachyon in bosonic string; usually this permits us to obtain results valid for superstring,
but the calculations are less complicated.)

 First of all, it is clear that string field theory can be treated in the framework of quantum mechanics, therefore it is possible to construct  an algebra of observables. Some of these observables  should be considered as quasilocal observables, however, it  is not quite clear what is the right choice of the algebra of quasilocal observables. Ten-dimensional group of translations acts on the algebra of observables.  It seems that
the asymptotic commutativity condition in  weak form is satisfied in the ground state representation
(at least in the framework of perturbation theory).

Let us discuss  the asymptotic commutativity in the
framework of light-cone bosonic open string field  theory. First of all
one should notice that for the free string the string field is in some sense local. Namely,
the commutator of string field with a string field
shifted by vector $\ba$ in spatial direction vanishes if $||\ba||$ is sufficiently large.  This follows from the fact [12],[13] that 
\begin{equation}
\label{lc}
[\Phi (X),\Phi (X')]=0 
\end{equation}
if 
\begin{equation}
\label{lcc}
2\delta x^-_0\delta x^+ -{\delta \bx _0}^2-2\sum _l(\delta \bx _l)^2_0<0
\end{equation}
where $\Phi (X)$ stands for the string field that depends on $x^+, x^-_0$ and transverse coordinates $$\bX (\sigma)=\bx _0+2\sum _l\bx _l\cos l\sigma.$$ (We use the notation $\delta X=X'-X.)$

The above statement  means that string fields commute outside the so-called string light cone
$\int d\sigma (X'(\sigma)-X(\sigma))^2 <0.$

It was shown in [14] that for interacting string (\ref {lc}) is violated. However, as was noticed in [15]   the commutator $[\Phi (X),\Phi (X')]$ is strongly suppressed for strings separated in spatial direction (at least in the first order of perturbation theory). Similar statements are true for closed strings and for superstrings. This indicates that we should expect asymptotic commutativity in the vacuum representation
of superstring theory. It  would be very interesting to prove this in the framework of perturbation theory.

We talked about light-cone string field theory, however, the situation in covariant string field theory should be similar. The case of non-interacting strings was analyzed in [16]; it was shown there that the commutator of string fields vanishes under the same conditions as in light-cone theory. Interacting strings where considered in [17] from a different viewpoint.
It was found in this paper that the theory can be written in a form when it is local in one light-like direction, but non-local in other directions  (this is similar to light-cone field theory). It is shown that the theory has a singular local limit and that close to this limit it is "almost local".  It is possible that the correct  algebra of observables
in covariant string field theory consists of operators constructed in [17]; it seems that these operators asymptotically commute (at least if we neglect the interaction of strings).

One can conjecture that compactifications of the superstring theory on Calabi-Yau 3-folds can be considered as ground states of the theory that are invariant with respect to four-dimensional subgroup of the translation group. Notice that Riemannian metric can be considered as a coherent state of gravitons. This means
that a compactification of string theory can be regarded as a state of string theory in flat space; in semiclassical approximation  this state is  a ground state.

It is important to emphasize that superstring theory induces gravity in four-dimensional space. The state of our world (expanding universe) is not a stationary state, therefore our considerations cannot be applied to this state. However, this is a state of local equilibrium. (Moreover, with great precision we can say
that the state of our world is locally a ground state.) We mentioned already that in this situation there exist local notions of space and time. This is precisely the situation of general relativity . (Recall that  Riemannian manifold can be described as
locally flat space.)

If this picture is correct then , probably, the string theory today is similar to quantum electrodynamics of early thirties. We should formulate the right mathematical problems and we should find a way to solve them. It seems
that the present paper paves the way to rigorous
formulation of problems of superstring field theory.
First of all one should specify the algebra playing the role of the algebra $\cA$ (of the algebra of quasilocal  observables). Time and space translations as well as Lorentz transformations should act on this algebra. Having the algebra and the action of translations we can give a definition of ground state. It is a difficult problem to find  ground states of the theory.
(Semiclassical considerations prompt a conjecture that along with ground state leading to ten-dimensional space-time superstring has ground states corresponding to
Calabi-Yau threefolds and leading to four-dimensional space-time.  It is possible, however,
that  not all states corresponding to Calabi-Yau manifolds are genuine ground states.)  Another hard problem is to find a particle spectrum in  given vacuum state. It is important to notice that
almost all particles of free string become unstable in interacting string theory.

 Of course, the main difficulty lies in finding the solutions, not in the formulation of problems.
In particular, we should go beyond the perturbation theory to 
prove equivalence of all existing superstring theories. 
However, finding an appropriate mathematical framework is a necessary step on the way to a complete understanding of string theory.

{\bf Acknowledgments}

I appreciate useful comments of D. Buchholz, M. Douglas, T.Erler, D. Gross, V. Kac,
M. Kontsevich, N. Moeller, N. Seiberg, B. Zwiebach.
 
{\bf References}

[1] V. Fateev,  A. Schwarz, On axiomatic scattering theory
Teor. Mat. Fiz., 14 (1973) 152-169, English translation in
Theor. Math. Phys,14 (1973) 112-124

[2] R. Haag, Local quantum physics, Springer, 1992

 [3] D. Ruelle, On the asymptotic condition in quantum field theory, Helv. Phys. Acta 35  (1962) 147
 
 [4] R. Haag, Quantum field theories with composite particles and asymptotic conditions, Phys. Rev., 112 (1958) 669
 
 [5] D. Buchholz, Harmonic analysis of local operators,
 Commun. Math. Phys. 129 (1990) 631-641
 
  [6] D.Buchholz, M. Porrmann, U. Stein, Dirac versus Wigner. Towards a universal particle concept in local quantum field theory. Phys. Lett. B267 (1991) 377
 
 [7] D. Buchholz,  On the manifestations of particles,
 hep-th/9511023
 
 [8]  M. Porrmann ,
  Particle Weights and their Disintegration I, II
    Commun.Math.Phys. 248 (2004) 269-304, 248 (2004) 305-333
    
[9]  C. A. Barata,PhD thesis, 1989;  S-matrix Elements in Euclidean Lattice Theories.
Rev. Math. Phys. 6, 497-513 (1994). 

[10] D. Buchholz, S.J. Summers, Scattering in relativistic quantum field theory:fundamental concepts and tools, math-ph/0509047
  
  [11] H. Araki, R. Haag,  Collision cross sections in terms of local observables,
Communications in Mathematical Physics,4 (1967)	77-91
  
  [12] E. Martinec,  Strings and Causality,
  hep-th/9311129
  
  [13] D. A. Lowe, Causal properties of free string field theory, hep-th/9312107
  
  [14] D. A. Lowe, L. Susskind, J. Uglum, Information Spreading in Interacting String Field Theory, hep-th/9402136

  [15] D.A. Lowe, J. Polchinski, L. Susskind, L. Thorlacius, J. Uglum, Black hole complementarity vs. Locality, hep-th/9506138
  
  [16] H. Hata, H. Oda, Causality in Covariant String Field Theory, hep-th/9608128

  [17] T. Erler, D. Gross, Locality, Causality, and an Initial Value Formulation for Open String Field Theory, hep-th/0406199  
  
  [18] 
D. Buchholz ,Current trends in axiomatic quantum field theory, hep-th/9811233 
 
\end{document}